# Design of a broadband and polarization insensitive THz absorber based on two layers of periodic arrays of graphene disks

Omid Mohsen Daraei[1], Pejman Rezaei[2,*], Seyed Amin Khatami[2], Pouria Zamzam[2], Mohammad Eskandari[1], Mohammad Bemani[1]

[1]Department of Electrical and Computer Engineering, University of Tabriz, Tabriz, Iran
[2]Electrical and Computer Engineering Faculty, Semnan University, Semnan, Iran

ABSTRACT

In this paper, we analytically design a simple configuration of a broadband THz and polarization-insensitive absorber. The mentioned absorber consists of two layers of graphene disks, and the transmission line model is considered for the whole of the proposed absorber's structure to design it accurately. Therefore, the input admittance of the designed absorber is obtained by the transmission line model. Also, the real part of the input admittance is approximately tuned to be matched to the free space admittance. In contrast, the imaginary part of it is closely adjusted to zero around the central frequency of the THz absorber. Using only just two layers of Periodic Arrays of Graphene Disks (PAGDs) with one kind of dielectric as the material of substrates, it causes that the absorption of the structure can be achieved higher than 90% by the Finite Element Method (FEM). Normalized bandwidth has reached up to 75.4% in 5 THz as the central frequency of the device. As the next step, we use the CST studio software to validate our designed absorber, and it will show that the numerical results will have the best matching with the analytical method.

## 1. Introduction

THz waves have many applications in biology and medicine [1], medical imaging [2], security [3], fingerprint [4], and so on. Scientists have designed and simulated many devices in the THz region such as emitters [5], [6], detectors [7], absorbers [8-13], and so on. One of the essential THz devices is the absorber. Graphene is a two-dimensional and single layer material based on hexagonal structures from carbon atoms, and because of its special characteristics such as supporting surface plasmon polaritons (SPPs) and having more surface plasmons than nobel metals, has been utilized in THz absorbers [16-18]. Also, another essential feature of the graphene-based absorber in comparison with other absorbers is its tunable electrical conductivity by changing its Fermi levels (chemical potentials) that can operate in different ranges of THz frequencies [13,14]. In the last years, the design of THz absorbers based on graphene with various shapes have been reported such as square patches [15], ribbons [13], disks [11],[14] and cross-shaped arrays [8]. In the recent research papers [13], [14], a layer of Periodic Arrays of graphene disks (PAGDs) has been modeled by a series RLC circuit where R, L, and C are resistor, inductor, and capacitor, respectively. Other types of THz absorber are based on H-shaped all-silicon arrays [25], vanadium dioxide metamaterials [26-28], and gold-film pattern [29]. These types of the absorber, in comparison to graphene-based absorbers, have a narrow bandwidth.

In this work, we design an ultra-wideband THz absorber with two layers of PAGDs that has an excellent operation in comparison with previous related works [9,23] in high THz frequencies. For the design of this THz absorber, the transmission line model (TLM) [13,14] is developed where two layers of the PAGDs are modeled as circuit components. As a result of simulations, the bandwidth of 90% absorption approximately reaches 3.77 THz with the central frequency of 5 THz for both TM and TE modes of the oblique incident THz wave. As a result, the normalized bandwidth of 90% absorption is achieved by 75.4% for this proposed absorber. On the other side, its absorption parameters can be adjusted by tuning the fermi levels (chemical potentials). The proposed device has been simulated both numerically and analytically using CST microwave studio software and the TLM, respectively. The rest of this paper consists of three sections. Section two describes the structure of the proposed THz absorber and its equivalent circuit model based on the TLM. In section three, simulation results have been analyzed, and the final section concludes the paper.

## 2. Structure and circuit model of the device

Fig. 1 shows the perspective view of the unit cell structure of the proposed THz absorber. As shown in fig. 1, the absorber is designed by two layers of PAGDs in dark color and two substrates made from the same material. As depicted in Fig. 1, a golden sheet in yellow color with the thickness of 1 μm that is used in the bottom of the structure, it has the role of metallic ground, so the incident THz wave cannot transmit through it because of the skin depth theory. In other words, the golden sheet acts as a short circuit that can reflect the incident THz wave thoroughly. The first layer of the graphene disks and the golden sheet are separated by a TOPAS polymer (cyclic olefin copolymers) with a little impurity which has the refractive index $n_{s1}$=1.5 in THz band. And also, the material of the second substrate is TOPAS polymer, which is located between two PAGDs. The thickness of graphene disks is assumed $t_g$ = 1 nm, which is 10 times greater than the thickness of a graphene layer. In order to have maximum absorption and minimum reflection, admittance matching to free space has been used, so based on the admittance matching, calculation of the first substrate thickness can be done by $H_1 = c_0 / 4n_1 f_0$, where $c_0$ is the speed of light in the free space, $n_1$ and $n_2$ are referred to refractive indices for the first and second substrate correspondingly while in this paper $n_1$ and $n_2$ are same, and $f_0$ is the central THz frequency of the absorber.

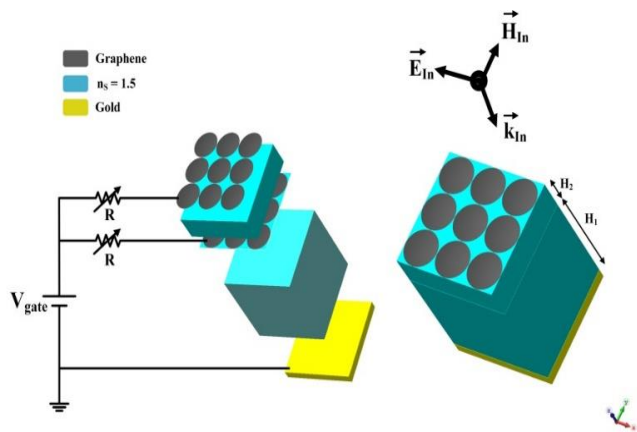

Fig. 1. The perspective view of the unit cell of the proposed absorber consists of one type of material for the first and second substrates and two layers of PAGDs.

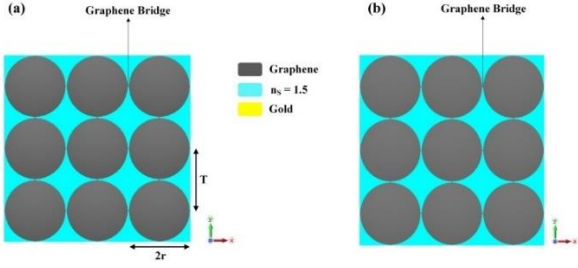

Fig. 2. Top views of the unit cell structure of the proposed absorber for (a) the first layer of PAGDs and (b) the second layer of PAGDs.

The surface conductivity of graphene contains intra-band and inter-band parts ($\sigma_{gr} = \sigma_{intra} + \sigma_{inter}$), and it can be calculated by Eq. 1. Because of the consideration of $E_f >> K_B T$ for THz frequencies, the main part in the surface conductivity is the intra-band section part so that the inter-band portion can be neglected in Kubo formula [19]. It should be noted that $E_F$, $\omega$, and $\hbar$ are the fermi level (chemical potential), the frequency in the THz regime, and the Planck's constant, respectively. Therefore, the Kubo formula can be simplified as Eq. 2.

$$\sigma_{gr} = \frac{e^2 K_B T}{\pi \hbar^2 \left( \tau^{-1} + j\omega \right)} 2\ln\left[ 2\cosh\left( \frac{E_F}{2K_B T} \right) \right] + \frac{e^2}{j4\pi\hbar} \ln\left[ \frac{2E_F - \hbar(\omega - j\tau^{-1})}{2E_F + \hbar(\omega - j\tau^{-1})} \right] \quad (1)$$

$$\sigma_{gr} = \frac{e^2 K_B T}{\pi \hbar^2 \left( \tau^{-1} + j\omega \right)} 2\ln\left[ 2\cosh\left( \frac{E_F}{2K_B T} \right) \right] \quad (2)$$

Where $K_B$, $e$, $T$, $\omega$, and $\tau$ refer to the Boltzmann constant, the charge of the electron, the temperature of the surrounding, the angular frequency, and the relaxation time of electrons, respectively. In this paper, the value of T is set to 300 K (room temperature), and $\tau$ will be calculated from the circuit model. So, we show the real part and the imaginary part of graphene conductivity at various Fermi levels in fig. 3. In the circuit model of PAGDs [13], a series RLC branch model has been proposed for each layer of graphene disks with periodic configuration, and it is depicted in fig. 4, the admittances of modeled series RLC branches for the first and second layers of PAGDs are illustrated by $Y_g$. In order to design a THz absorber based on the PAGDs, the TLM has been developed for the whole structure of the absorber, in this regard the golden sheet is considered as a short circuit, so its admittance will be $Y_{Au} \approx \infty$, and also the admittance of dielectric substrate can be calculated by $Y_s = Y_0 n_s$ in the TLM.

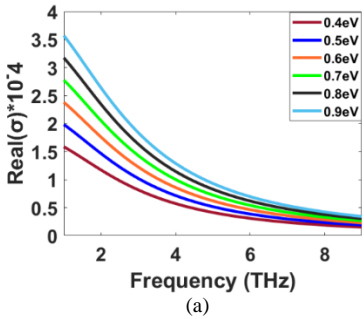

(a)

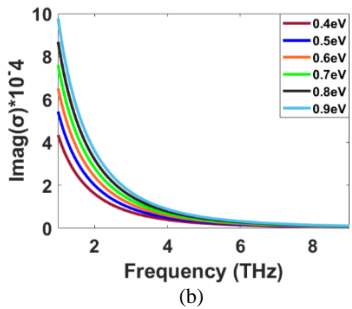

(b)

Fig. 3. At different Fermi levels of graphene. (a) the real part and (b) the imaginary part of the frequency as a function of graphene conductivity.

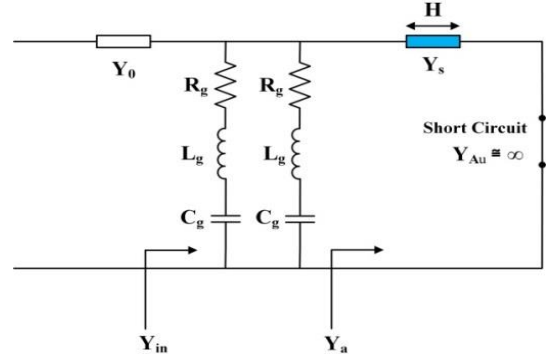

Fig. 4. The transmission line model of the mentioned THz absorber.

For the normal THz incident wave, input admittances for the TLM of the proposed absorber can be calculated as:

$$Y_a = \frac{Y_s}{j\tan(k_s H)} = -jY_s \cot(k_s H) \quad (3)$$

$$Y_g = \frac{1}{R_g + j(\omega L_g - \frac{1}{\omega C_g})} \quad (4)$$

$$Y_{gp} = Y_g + Y_g = 2Y_g \quad (5)$$

$$Y_{in} = Y_{gp} + Y_a = \frac{2}{R_g + j(\omega L_g - \frac{1}{\omega C_g})} - jY_s \cot(k_s H) \quad (6)$$

Where $H$ and $K_s$ refer to the thicknesses of the substrates and the propagation constants of the incident THz wave, respectively. There is a direct relationship between the input admittance of the mentioned absorber and the thicknesses of the substrates. So, for achieving the wideband absorption, the real part of the input admittance should be approximately adjusted to the free space admittance ($Y_0$), and the imaginary part of it should be infinite, so we have:

$$\text{Re}(Y_{in})\big|_{f=f_0} = \frac{1}{R_{gp}} = \frac{2}{R_g} = \alpha Y_0$$

$$\text{Im}(Y_{in}) = \text{Im}(Y_{gp}) + \text{Im}(Y_a)$$

Where $\alpha$ is a numerical parameter larger than unity, and the minimum allowed absorption determines its maximum value. From the transmission line model, the absorption in the central frequency can be expressed in terms of $\alpha$: $A = 1 - \left[ (1-\alpha)/(1+\alpha) \right]^2$. Therefore, we set $\alpha = 1.702$ throughout this work to ensure absorption larger than 90%. According to the approximately adjusted input admittance of the equivalent transmission line model to the free space admittance, the quantities of $R_g$, $L_g$ and $C_g$ can be determined, so these quantities can be considered as input parameters for calculating the radii and periods of the PAGDs for the proposed absorber. Moreover, the equivalent resistance, capacitance, inductance, and relaxation time for each layer of the PAGDs have been calculated [14] when they are utilized in the $n^{th}$ mode as follows:

$$R_{g_n} = \frac{K_n T^2 (\frac{h}{2\pi})^2}{S_n^2 e^2 \pi E_F \tau_n} \quad (9)$$

$$L_{g_n} = \frac{K_n T^2 (\frac{h}{2\pi})^2}{S_n^2 e^2 \pi E_F} \quad (10)$$

$$C_{g_n} = \frac{\varepsilon_0 \pi^2 S_n^2 n_s^2}{T^2 K_n q_{11}} \quad (11)$$

$$\tau_n = \frac{L_{g_n}}{R_{g_n}} \quad (12)$$

The values of $e$ (the electron charge) and $h$ (Planck's constant) are equal to $e \cong 1.602176 \times 10^{-19} [c]$ and, $h = 6.626 \times 10^{-34} [J.s]$. In Eq. (11), $n_s$ is the refractive index of the substrate, which are surrounding the graphene disks, and $S_n$ could be derived from the integral of the eigenfunctions in [14]. Also, in Eq. (11), $q_{11}$ shows the first eigenvalue of the equation that controls the surface current density on the graphene disks. Hence, the values of $q_{11}$ for various 2r/T are illustrated in table 1 [14]. By considering only the first mode of the resonance frequency of PAGDs in this paper, the parameters of the first mode will be $S_1 = 0.6087r$ and $K_1 = 1.2937$ [14]. According to the quantities of table 1, the quantity of $q_{11}r$ for 2r/T = 0.98 is considered approximately 0.39, so it can be determined as follow $q_{11}r = 0.3873$ [14]. The radius of graphene disks is r = 1.5547 μm. The period of the first and second layers of graphene disks is $T$. So, based on the equivalent Eqs. (9) – (11) and quantities of table 1, the parameters of the TLM can relate to the parameters of the practical method, so the quantities of the radii and periods of the first and second layers of the PAGDs can be calculated by the above-mentioned equations.

Table. 1. The calculated Eigenvalues of the equation for controlling the first mode of surface current density on graphene disks [14].

| 2r/T | 0.1 | 0.5 | 0.9 | 0.98 |
|---|---|---|---|---|
| q₁₁r | 0.539 | 0.527 | 0.417 | 0.3873 |

Furthermore, the angular resonance frequencies of the first and second layers of the PAGDs can be evaluated by $\omega_{0g1}$ and $\omega_{0g2}$, respectively, and they can be written as:

$$\omega_{0g} = \frac{1}{\sqrt{L_g C_g}} \quad (13)$$

### 3. Simulation Results

In this section, a THz absorber based on two layers of the PAGDs has been designed and simulated. The operation of the broadband THz absorber based on two layers of the PAGDs has been studied by developing the TLM. In this model, the real part of the input admittance has been tuned to be closely matched to $Y_0$ (the free space admittance), and also the imaginary part of it could be adjusted to infinite around the central frequency of the absorber. The calculated parameters for designing the proposed device have been listed in table 2. For the 3-D simulation of the proposed structure, the CST software MW studio has been used. The parameters of 3-D simulations are described as follows. To calculate the $H$, the input admittance ($Y_{in}$) of the absorber is calculated.

Table. 2. The calculated values of parameters for designing the proposed THz absorber.

| | The 1st Layer of PAGDs | The 2nd Layer of PAGDs |
|---|---|---|
| Proposed THz Absorber | T = 3.1736 μm | T = 3.1736 μm |
| | r = 1.5547 μm | r = 1.5547 μm |
| | H₁ = 10 μm | H₂ = 0.2 μm |

The spectra of the normalized input admittance of this THz absorber with the real and imaginary parts are depicted in Fig. 5.

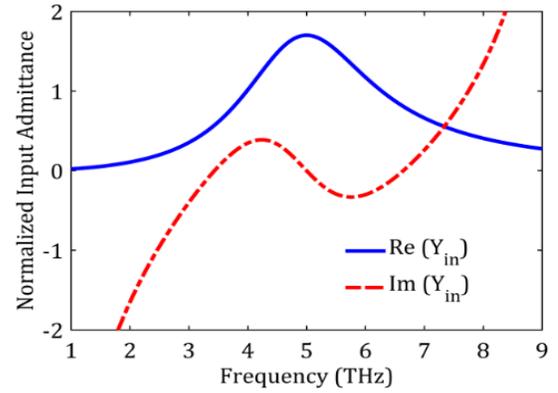

Fig. 5. The spectra of the real and imaginary parts of the normalized input admittance, achieved from the TLM.

As seen the differential of the imaginary part of input admittance is smaller than zero in the central frequency of 5 THz, and we can write:

$$\frac{d}{df}\text{Im}(Y_{in})\Big|_{f=f_0} = \frac{d}{df}\text{Im}(Y_{gp})\Big|_{f=f_0} + \frac{d}{df}\text{Im}(Y_a)\Big|_{f=f_0} < 0 \quad (14)$$

On the other side, the imaginary part of $Y_{gp}$ and $Y_a$ around the central frequency can be calculated using the Tailor expansion. Assuming $K_s H \approx \pi/2$ and $Y_{Au} \to \infty$, it is obtained that:

$$\text{Im}(Y_a)\Big|_{f \approx f_0} \cong -Y_s \cot(k_s H) \cong Y_0 n_s \frac{\pi}{2}(\frac{f}{f_0} - 1) \quad (15)$$

So, by considering $LC = 1/(2\pi f_0)^2$, the admittance of PAGD for frequencies around $f_0$ can be expressed as:

$$\text{Im}(Y_{gp})\Big|_{f \approx f_0} \cong -\frac{4\pi L_g}{R_g^2 f}(f^2 - f_0^2) \quad (16)$$

By inserting Eq. 15 and 16 in Eq. 14, we have:

$$\frac{Y_0 n_s \pi}{2f_0} - \frac{8\pi L_g}{R_g^2} < 0 \quad (17)$$

If we consider the constant of $\beta'$, we can calculate the $\tau_g$ as fallow:

$$\frac{Y_0 n_s \pi}{2f_0} - \beta' \frac{8\pi L_g}{R_g^2} = 0 \to \frac{n_s \pi}{2Z_0 f_0} = \beta' \frac{8\pi L_g}{R_g^2} \xrightarrow{\frac{L_g}{R_g} = \tau_g} \tau_g = \frac{n_s}{16\alpha\beta f_0} \quad (18)$$

It can be seen that $\tau_g$ is proportional to the central frequency ($f_0$) and can be tuned by the Fermi energy. Finally, we have the $E_F$ as fallow:

$$E_F = \frac{\tau_g e v_F^2}{\mu} \quad (19)$$

In which $v_f$ is the Fermi velocity and $\mu$ is the mobility of electrons. The fill factor of the graphene array disks, 2r/T, is then obtained from Eq. 9:

$$\frac{2r}{T} = \sqrt{\frac{K_1 \hbar^2}{0.093\pi E_F \tau_g R_g e^2}} \quad (20)$$

$$r = \frac{a_1 E_F \, e^2}{\pi \varepsilon_{eff} \hbar^2 \omega_0^2} \quad (21)$$

$$a_1 = r q_{11} \quad (22)$$

Where $r$ is the radius of disks. In Fig. 6, the complete design flow of the analytical method.

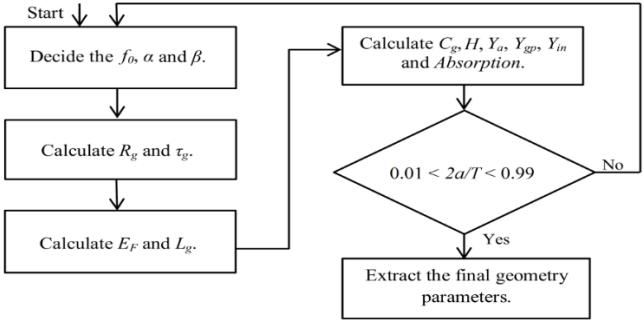

Fig.6. The complete design flow of the analytical method.

In the next step, the accuracy of the transmission line model and the simulation are compared. The calculated spectrum of the ultra-wideband absorption by the FEM (Finite Element Method) and the calculated TLM are depicted in Fig. 7(a). As shown, the bandwidth of the absorber is about 3.77 THz and provides an ultra-wideband absorber. As illustrated in Fig. 7(b), the obtained return loss coefficient spectra of this structure are less than 0.1, as shown by the FEM and the TLM. As we know, the return loss coefficient can be calculated by $\Gamma = S_{11} = (Z_{in} - Z_0)/(Z_{in} + Z_0)$. Because of the fact that the incident THz wave that cannot transmit through the golden layer, the transmission coefficient ($S_{21}$) can be neglected in the absorption relationship so that it can be written as $A = 1 - |S_{11}|^2$.

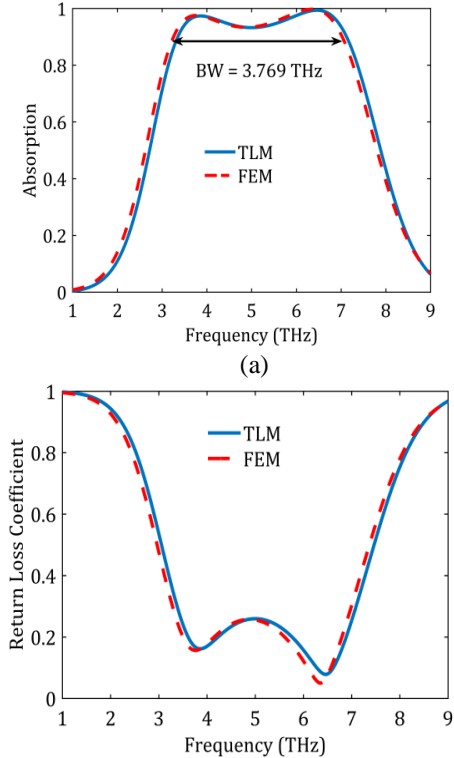

Fig. 7. (a) The absorption and (b) return loss coefficient spectra of the designed device, calculated by the FEM and TLM.

In this work, we consider the incident THz wave with both transverse magnetic (TM) and transverse electric (TE) polarizations with incident angle $\cong 0$. Due to the disk shape of graphene layers, the incident wave with any polarization excites surface plasmons; as a result, surface plasmon polaritons (SPPs) waves propagate in both directions x, y. Because of the propagation of the SPPs, the incident wave can be absorbed. In fig. 8 (a), the absorption spectra for both TM and TE modes are demonstrated based on these considered Fermi energies for the first and second layers of the PAGDs $E_{F1}$ = 0.485 eV and $E_{F2}$ = 0.485 eV. And also, in fig. 8(b), the return loss coefficient spectra for these modes are determined. There is no any difference between TE and TM modes and the absorber is polarization insensitive.

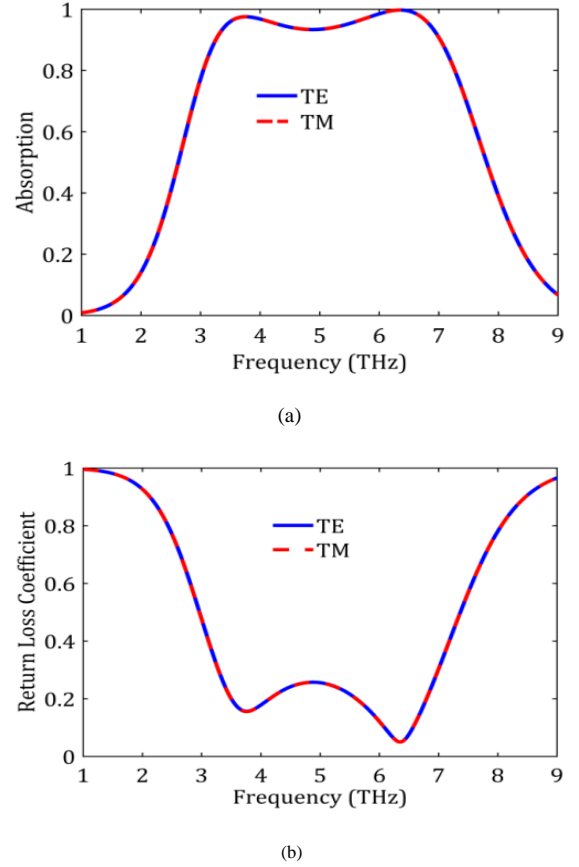

(a)

(b)
Fig. 8. (a) The absorption and (b) return loss coefficient spectra of the proposed absorber for TM and TE polarizations of the incident THz wave, calculated by the FEM.

The values of the absorption and return loss coefficients can be controlled by the fermi levels (chemical potentials). The sensitivity of the absorber with respect to variations in the Fermi energy is also examined in fig. 9. It is evident from this figure that the absorption and return loss coefficient spectra are not changed significantly for ±10% variation in the Fermi energy.

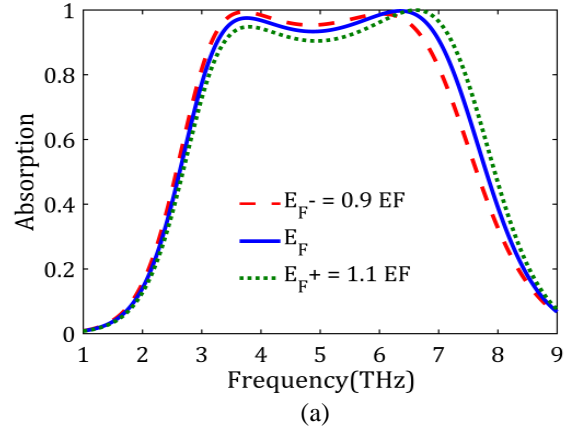

(a)

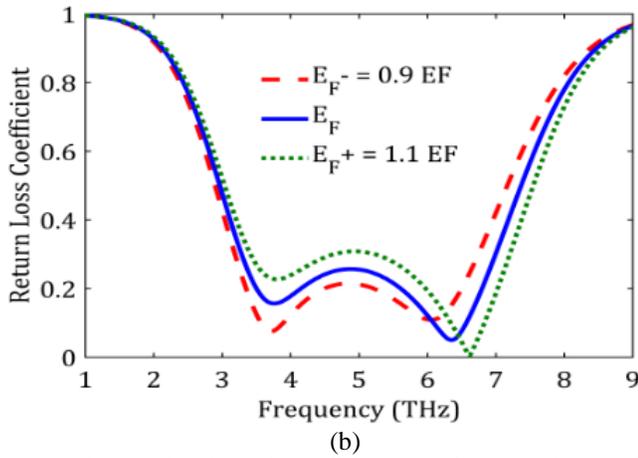

(b)

Fig. 9. (a) Absorption (b) and return loss coefficient spectra of the proposed absorber for various $E_F$, obtained by FIT.

In fig. 10, the absorption and return loss coefficient spectra's dependency to the incident angle of the electromagnetic wave from 0 to 60 degrees are depicted. It is seen that for angles of 0 and 20 degrees, the absorption spectra are the same, but for the angle of 60 degrees, the spectrum is being narrower but still is a perfect absorber in some frequencies.

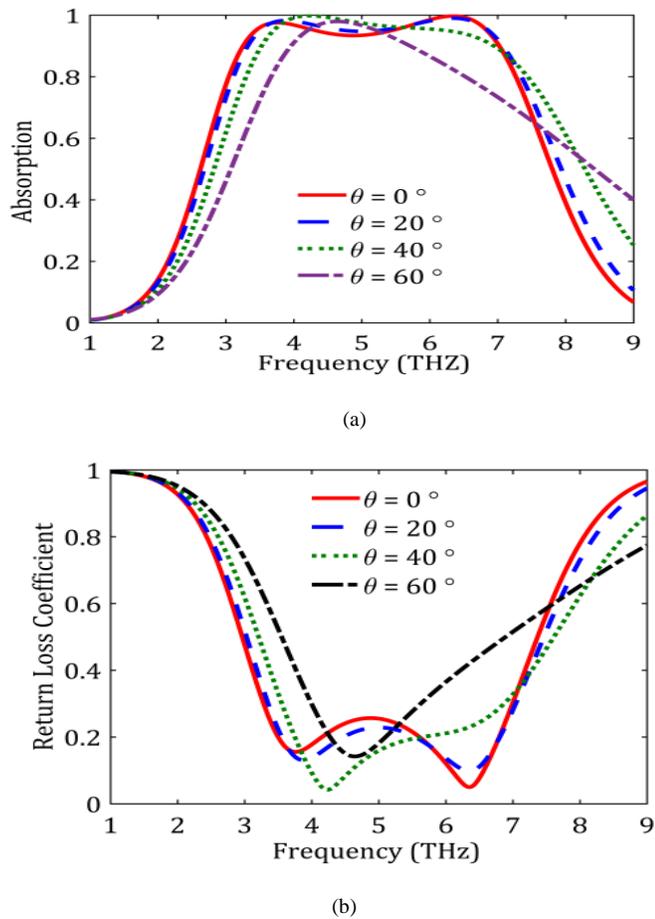

Fig. 10. The angle dependence of absorption and return loss coefficient spectra.

The surface current density on the graphene layers for both TE and TM modes in the frequency of $f_0 = 1$ THz (frequency related to the lowest absorption), $f_0 = 5$ THz (central frequency) and $f_0 = 6.35$ THz (frequency related to the highest absorption) have been shown in fig. 11. It is seen that the current density in the first layer of graphene is higher than the second one and it can be concluded that the first layer plays main role in absorbing of the waves. Also, in this figure, the excitation of plasmons for both modes is depicted. As a result of these excitations, localized surface plasmons (LSPs) with frequencies at the range of absorber frequencies excited in the graphene discs. Surface plasmons are the oscillations of electric charges of the conductor formed in the conductor-insulator interface and can be stimulated by the electric field of the incident light [15-17]. Strong electric fields created by surface plasmon excitements enhance the absorption, which is proportional to the square of the electric field.

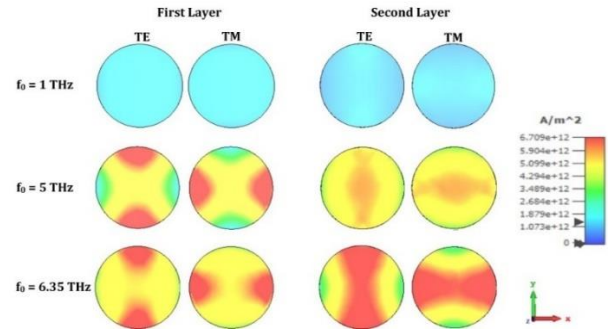

Fig. 11. The distribution of surface current density for TE and TM modes for first and second layers of PAGDs in $f_0 = 1$ THz, $f_0 = 5$ THz and $f_0 = 6.35$ THz.

In Fig. 12, the profile of the electric field in the absorber for $f_0 = 1$ THz (frequency related to the lowest absorption), $f_0 = 5$ THz (central frequency) and $f_0 = 6.35$ THz (frequency related to the highest absorption) have been shown. It is seen that the distribution of electric field between the layers of disks at $f_0 = 6.35$ THz is more significant and it results in absorption enhancement. This is while at $f_0 = 1$ THz, the intensity of the electric field is low and consequently makes the absorption to be low. The phenomenon behind the absorption enhancement is related to the excitation of plasmons in the lateral sides of the Graphene disks, so they localize the electric field of the wave and trap it through the absorber.

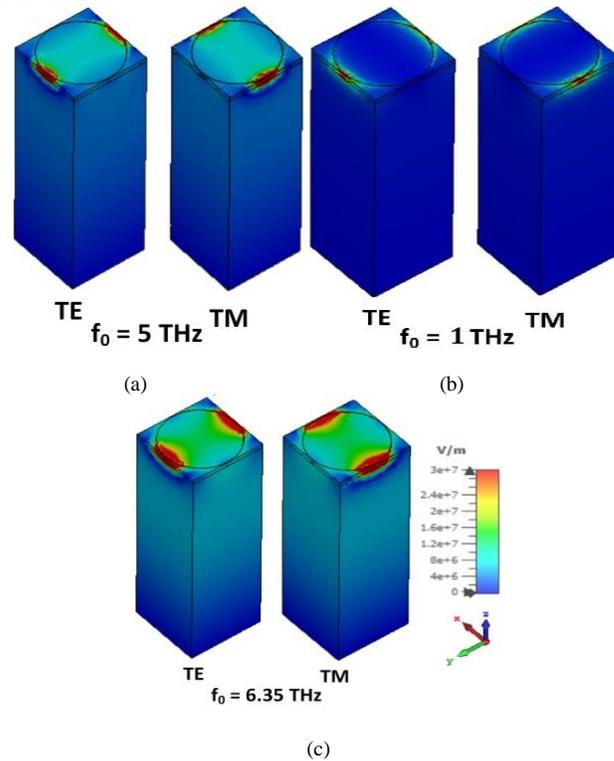

Fig. 12. The distribution of Electric field for TE and TM modes in $f_0 = 1$ THz, $f_0 = 5$ THz and $f_0 = 6.35$ THz.

As a result, the operation of our proposed ultra-wideband THz absorber and some of the best previous broadband and ultra-broadband absorbers are given in table 3. This work has some benefits than the other works such as high normalized bandwidth (BW/f0 (%)) equal to 75.4%, small size, insensitive to TE and TM polarizations, using only just two layers of PAGDs, and simple structure.

Table 3. The comparison of our proposed device operation with some of the previous THz absorbers.

| Reference | Number of Layers | Device Height(μm) | $f_0$ (THz) | BW/$f_0$ (%) | Polarization |
|---|---|---|---|---|---|
| [10] | 3 | 11 | 6.3 | 25.4 | TE & TM |
| [24] | 9 | 28.5 | 5.4 | 88.8 | TM |
| [31] | 5 | 26.5 | 1.93 | 65 | TE & TM |
| [32] | 3 | 20.5 | 1.94 | 81 | TE & TM |
| [33] | 5 | Not reported | 1.38 | 21 | TE & TM |
| [34] | 3 | 42.9 | 3.03 | 30 | TE |
| [35] | 3 | 40.5 | 2.56 | 67 & 85 | TE & TM |
| Proposed Device | 5 | 10.6 | 5 | 75.4 | TE & TM |

## 4. Conclusion

In this paper, we design analytically a simple configuration of a broadband THz and polarization insensitive absorber using two layers of the PAGDs. The RLC branch has been modeled for each layer of Graphene disks. Then, the transmission line model has been developed for the whole of the absorber configuration, in which the mentioned series RLC branch was considered. And also, the TLM could be used for achieving the value of the input admittance of the designed absorber. Also, the thickness of graphene disks was assumed with 1 nm in this work that is approximately 10 times greater than the thickness of one graphene layer. Finally, based on the simulations, the obtained bandwidth of the proposed broadband THz absorber equals to 3.77 THz with 5 THz of the central frequency by FEM and TLM, in which the return loss coefficient was less than 10% in a THz regime. On the other side, both analytical (TLM method) and numerical (FEM did by CST) results showed the best fitting.

## References

1. P. H. Siegel, Terahertz technology in biology and medicine, IEEE Trans Microw Theory Tech **52** (10) (2004) 2438-2447.
2. L. Dwight, R. William, M. Shur, Terahertz Sensing Technology: Electronic devices and advanced systems technology. (1) (2003).
3. F. John, B. Schulkin, F. Huang, D. Gary, R. Barat, F. Oliveira, and D. Zimdars, THz imaging and sensing for security applications—explosives, weapons and drugs, Semiconductor Science and Technology **20** (7) (2005) S266.
4. K. Kawase, Y. Ogawa, Y. Watanabe, and H. Inoue, Non-destructive terahertz imaging of illicit drugs using spectral fingerprints, Optics express **11** (20) (2003) 2549-2554.
5. K. Goudarzi, S. Matloub, and A. Rostami, Generation of two-color terahertz radiation using Smith–Purcell emitter and periodic dielectric layer, Opt Quantum Electron **51** (2) (2019) 48.
6. W. Knap, S. Rumyantsev, M. Vitiello, D. Coquillat, S. Blin, N. Dyakonova, and T. Nagatsuma. Nanometer size field effect transistors for terahertz detectors, Nanotechnology **24** (21) 2013 214002.
7. M.Mittendorff, et al. Ultrafast graphene-based broadband THz detector, Applied Physics Letters **103**(2) (2013) 021113.
8. S. Ke,B.Wang, H. Huang, H. Long, K. Wang, and P. Lu, Plasmonic absorption enhancement in periodic cross-shaped graphene arrays, Optics express **23** (7) (2015) 8888-8900.
9. P. Zamzam, P. Rezaei, and S. A. Khatami, Quad-band polarization-insensitive metamaterial perfect absorber based on bilayer graphene metasurface,. Physica E Low Dimens. Syst. Nanostruct 128 (2021) 114621.
10. P. Zamzam and P. Rezaei, A terahertz dual-band metamaterial perfect absorber based on metal-dielectric-metal multi-layer columns, Opt Quantum Electron 53 (2) (2021) 1-9.
11. A. Norouzi and P. Rezaei, Broadband polarization insensitive and tunable terahertz metamaterial perfect absorber based on the graphene disk and square ribbon, Superlattice microst (2022) 107153.
12. L.Ye, Y.Chen, G.Cai, N.Liu, J.Zhu, Z.Song and Q.H.Liu, Broadband absorber with periodically sinusoidally-patterned graphene layer in terahertz range, Optics express **25** (10) (2017) 11223-11232.
13. A.Khavasi, Design of ultra-broadband graphene absorber using circuit theory, *JOSA B* **32**(9) (2015) 1941-1946.
14. S. Barzegar-Parizi, B. Rejaei,and A. Khavasi, Analytical circuit model for periodic arrays of graphene disks, IEEE Journal of Quantum Electronics **51** (9) (2015) 1-7.
15. Hu, Dan, et al. Tunable broadband terahertz absorber using a single-layer square graphene patch with fourfold rotationally symmetric groove, *Optical Materials* 109 (2020) 110235.
16. M. Kuisma, A. Sakko, T. P. Rossi, A. H. Larsen, J. Enkovaara, L. Lehtovaara, and T. Rantala, Localized surface plasmon resonance in silver nanoparticles: Atomistic first-principles time-dependent density-functional theory calculations, Physical Review B **91**(11) (2015) 115431.
17. E. Petryayeva, and U. J. Krull, Localized surface plasmon resonance: Nanostructures, bioassays and biosensing—A review Analytica chimica acta **706** (1) (2011) 8-24.
18. Y. Chen, and H. Ming, Review of surface plasmon resonance and localized surface plasmon resonance sensor, Photonic Sensors **2** (1) (2012) 37-49.
19. G. Shen, M. Zhang, Y. Ji, W. Huang, H. Yu and J. Shi, Broadband terahertz metamaterial absorber based on simple multi-ring structures, AIP Advances **8** (7) (2018) 075206.
20. K. Arik, S. AbdollahRamezani, and A. Khavasi, Polarization insensitive and broadband terahertz absorber using graphene disks, Plasmonics **12** (2) (2017) 393-398.
21. S. Wu, D. Zha, L. Miao, Y. He, and J. Jiang, Graphene-based single-layer elliptical pattern metamaterial absorber for adjustable broadband absorption in terahertz range, Phys. Scr. 94 (2019) 105507.
22. C. Iman, B. Sadegh, Ultra-broadband terahertz absorber based on graphene ribbons, Optik **172** (2018) 1026-1033.
23. G. W.Hanson, Dyadic Green's functions for an anisotropic, non-local model of biased graphene, IEEE Transactions on antennas and propagation **56** (3) (2008) 747-757.
24. S. Biabanifard, S. Asgari, S. Asadi and C.E. Mustapha, Tunable ultra-wideband terahertz absorber based on graphene disks and ribbons, Optics Communications **427** (2018) 418-425.
25. Z. Xu, D. Wu, Y. Liu, C. Liu, Z. Yu, L. Yu, and H. Ye, Design of a tunable ultra-broadband terahertz absorber based on multiple layers of graphene ribbons, Nanoscale research letters **13** (1) (2018) 1-8.
26. X Zhao, Y. Wang, J. Schalch, G. Duan, K. Cremin, J. Zhang, and X. Zhang, Optically modulated ultra-broadband all-silicon metamaterial terahertz absorbers, *ACS Photonics* **6** (4) (2019) 830-837.
27. Z. Song, K. Wang, J. Li,, and Q. H. Liu, Broadband tunable terahertz absorber based on vanadium dioxide metamaterials, Optics Express, **26** (6) (2018) 7148.
28. Z. Song, K. Wang, J. Li, and Q. H. Liu, Broadband tunable terahertz absorber based on vanadium dioxide metamaterials, Optics express, **26** (6) (2018) 7148-7154.
29. H. Liu, Z. Wang, L. Li, X. Fan, and Y. Tao, Vanadium dioxide-assisted broadband tunable terahertz metamaterial absorber, Scientific reports. **9** (1) (2019) 1-10.
30. Y. Cheng, H. Chen, J. Zhao, X. Mao, and Z. Cheng, Chiral metamaterial absorber with high selectivity for terahertz circular polarization waves, Optical Materials Express. **8** (5) (2018) 1399-1409.
31. L. Ye, Y. Chen, G. Cai, N. Liu, J. Zhu, Z. Song, Q.H. Liu, Broadband absorber with periodically sinusoidally-patterned graphene layer in terahertz range, Opt Express 25 (10) (2017) 11223–11232.
32. Y. Li, J. Wu, C. Wang, et al., Tunable broadband metamaterial absorber with single-layered graphene arrays of rings and discs in terahertz range, Phys. Scripta 94 (3) (2019), 035703
33. J. Zhu, C. Wu, Y. Ren, Broadband terahertz metamaterial absorber based on graphene resonators with perfect absorption, Results Phys. 26 (2021) 104466.
34. A. Mohanty, O.P. Acharya, B. Appasani, et al., A broadband polarization insensitive metamaterial absorber using Petal-shaped structure, Plasmonics 15 (2020) 2147–2152
35. A. Norouzi Razani, P. Rezaei, Broadband polarization insensitive and tunable terahertz metamaterial perfect absorber based on the graphene disk and square ribbon, Superlattice and microst. 107153, 2022.